\documentclass[aps,prper,twocolumn,floats,floatfix,longbibliography,superscriptaddress,a4paper]{revtex4-1}

\usepackage[T1]{fontenc}
\usepackage[utf8]{inputenc}
\usepackage{graphicx}
\usepackage{color}
\usepackage{enumitem}
\usepackage[cmex10]{amsmath}

\begin{document}


\title{Student Conceptual Difficulties in Hydrodynamics}


\author{Alvaro \surname{Suarez}}
\affiliation{CES-ANEP, Uruguay}

\author{Sandra \surname{Kahan}}
\affiliation{Instituto  de F\'\i sica,  
             Facultad de Ingenieria,
Universidad de la Rep\'ublica, Uruguay}

\author{Genaro \surname{Zavala}}
\affiliation{Tecnologico de Monterrey, School of Engineering and Sciences, Monterrey, Mexico}

\affiliation{Universidad Andres Bello, School of Engineering, Santiago, Chile}

\author{Arturo C. \surname{Marti}}
\affiliation{Instituto  de F\'\i sica,  
             Facultad de Ciencias,
Universidad de la Rep\'ublica, Uruguay}


\date{\today}

\begin{abstract}
We  describe a study on the conceptual difficulties faced by college students in understanding hydrodynamics of ideal fluids.  This
study was based on responses obtained in hundreds of written exams and oral interviews, which were held with first-year Engineering and
Science university students.  Their responses allowed us to identify a series of misconceptions unreported in the literature so far. 
The study findings demonstrate that the most important difficulties arise from the students' inability to establish a  link 
between the kinematics and dynamics of moving fluids, and from a lack of understanding regarding how different regions of a system interact.
\end{abstract}


\maketitle

\section{INTRODUCTION}

The physics of ideal fluids (non-viscous and incompressible) is studied at the introductory level in the 
first-year Physics and Engineering university courses, as well as in those related to Medicine and Life Sciences. 
An in-depth understanding of this topic requires, in addition to a knowledge of the basics of classical mechanics (statics,
kinematics and dynamics), knowledge of the specific concepts to fluids such as current lines, 
pressure, propulsion, and conservation of different physical magnitudes.

Physics education research shows that the conceptual difficulties to understand the phenomena associated with fluids have
received relatively uneven attention. On the one hand, the difficulties associated with hydrostatic principles have been deeply
analyzed by various authors, who showed how students continue to present serious difficulties in understanding the basics, even after
attending university courses where these topics are covered
(see \cite{besson2004students,buteler2014aprendizaje,goszewski2013exploring,loverude2003helping,loverude2010identifying,melo2016learning}
among others). On the other hand, students' understanding   of ideal fluid hydrodynamics has received less
attention. Studies in existing literature mainly relate to the
application of Bernoulli's equation and the results that might derive
from it \cite{barbosa2013construccion,recktenwald2009simple,vega2017dificultades,brown2017engineering}.

As a consequence, there are many open questions regarding students’ understanding of ideal fluid hydrodynamics that need to be addressed. 
How do they interpret the origin of forces acting on a volume element of a moving fluid? Do they connect changes in velocity 
(i.e., magnitude and direction) with pressure gradients? How do they apply conservation of mass in contexts other than fluids confined in pipes? 
These and other pertinent questions were addressed in this study, 
which is based on the analysis of midterm tests and exams, and of the responses obtained in interviews conducted with students who successfully passed 
the course of General Physics, which covered fluid mechanics topics. 

This work is organized as follows: The next section discusses the context of this study and describes some of the most relevant previous research. 
Section III details the methodology and describes the population participating in the study. 
Section IV presents the results from a detailed analysis of the responses given by students in midterm tests and exams. 
Section V presents students’ interviews protocols, problems and their responses. Section VI discusses the results, while the last section presents the conclusions.

\section{PREVIOUS RESEARCH}

\subsection{Students' conceptions in hydrodynamics}

Some authors that address students’ concepts, in the context of ideal fluid hydrodynamics, 
focused specifically on the misconception that, the greater the fluid velocity the greater its pressure.
The typical case where this misconception can be found is the problem of a liquid with negligible
viscosity which flows through a narrow pipe, as indicated in Fig.~\ref{fig1}.
In this problem, as shown below, the correct application of the hydrodynamic laws leads to the conclusion
that the greater the velocity is, the lower the pressure will be. However, it seems to be a popular misconception 
which claims that the opposite situation occurs.

\begin{figure}[h]
\includegraphics[width=0.8\columnwidth]{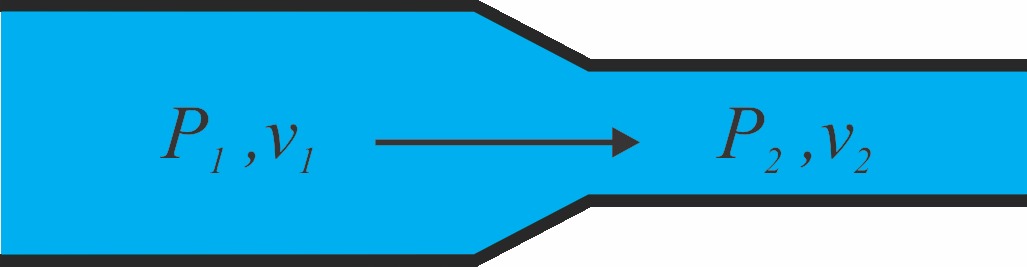}
\caption{
Flow along a narrowing pipe. $P_1$ and $v_1$ ($P_2$ and $v_2$) indicate the pressure and the velocity before (after) the narrowing.
\label{fig1}
}
\end{figure}

This error might have different origins. Martin \cite{martin1983misunderstanding} argued that the students' difficulties were caused 
by how they interpret the everyday experiment of covering a hose's outlet mouth to increase the exit velocity: 
the greater the force exerted by the water on the finger, the smaller the outlet must be.
Since students associate this force with the pressure of the fluid inside the hose, they understand that the higher the velocity,
the greater the pressure of the fluid in motion must be. According to Barbosa \cite{barbosa2013construccion}, the interpretation that allowed
students to conclude that the higher the pressure the greater the
velocity is based on the belief that pressure is equal to force, and to the fact that 
they link force to velocity rather than to acceleration. Brown \textit{et al.} \cite{brown2017engineering}
premised that this previous concept is related to the notion that liquids are
compressible, given that when the water flows from a wide pipe to a
narrower one, the water compresses, increasing the pressure. According to \cite{vega2017dificultades}, this concept is derived 
from previous interpretations of relating  pressure to the space occupied by the fluid \cite{besson2004students,goszewski2013exploring}, which
led to  the  assumption that pressure increases in narrower places, and, therefore, the velocity is greater.

This previous conception regarding the relationship between pressure and velocity was also demonstrated when implementing the Fluid Mechanics
Inventory Test \cite{watson2015refinement} as well as the Thermal and Transport Concept Inventory \cite{olds2004preliminary} with students attending
fluid mechanics courses. The results indicate that this concept persists even after students had successfully passed these courses.

\subsection{Bernoulli's equation in the context of ideal fluid hydrodynamics}

Bernoulli's equation is normally inferred in introductory textbooks \cite{halliday2001physics,young2008sears,tipler2007physics,serway2013physics}
through the work-kinetic energy theorem, applied to a confined and stationary ideal fluid. By linking
the kinetic and gravitational potential energy, per volume unit, with
the work of the pressures at the borders of the region of interest, we can see that the amount
\begin{equation}
P + \rho g h + \frac{\rho v^2}{2}
\end{equation}
is constant at any point of a streamline or a streamtube. When this result is
applied to a fluid flowing in a horizontal narrowing pipe, as shown in
Fig.~\ref{fig1}, the following occurs
\begin{equation}
P_1 +  \frac{\rho v_1^2}{2} = P_2 +  \frac{\rho v_2^2}{2}.
\end{equation}
Furthermore, with the continuity equation, the students easily deduce
that the greater the fluid velocity is, the lower the pressure will be. This is
textbook case application  of Bernoulli's equation.

We must emphasize that this relationship is not valid for every context; it is true only if the hypotheses under which it is deduced have been
verified. In particular, its application on different current lines or unconfined
flows can lead to incorrect predictions. This aspect has been discussed in numerous  studies, for example 
\cite{bauman1994interpretation,bedford1977misinterpretation,murphy1986bernoulli,eastwell2007bernoulli,kamela2007thinking,smith1972bernoulli,weltner2011misinterpretations}.

\begin{figure}
\includegraphics[width=0.350\columnwidth]{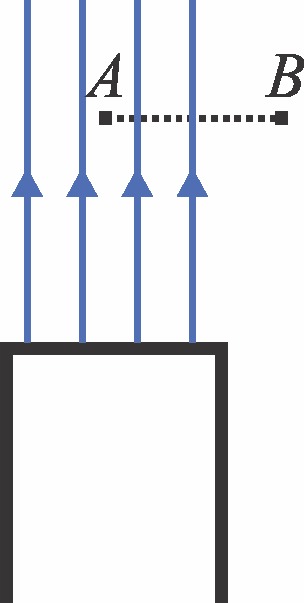}
\caption{Fluid jet immersed in a still medium}
\label{fig2}
\end{figure}

A notable example of an incorrect application is presented by  Kamela \cite{kamela2007thinking}, who discussed the pressures of a fluid
jet immersed in a resting environment. The pressure at a point within  the jet (point $A$ of Fig.~\ref{fig2}) is equal to the pressure at
a point outside of it (point B).
Using this simple situation and Newton's laws, Kamela demonstrated that the premise of higher speed-lower pressure is
not always valid. Indeed, 
if the pressure at point $A$ was lower than at point B, the pressure gradient accelerates the fluid along the line that 
joins these points.  The flow would not be uniform, and the current lines could not be parallel. It can then be concluded that the pressures 
at point $A$ and B must be equal. Obviously, in this example, Bernoulli’s equation cannot be applied, since there is no  streamline through point $B$.

This type of dynamic analysis, based on the measurement of a pressure gradient, 
is not commonly found in textbooks, which are usually focused on the consequences of applying both the continuity and 
Bernoulli’s equations without discussing in detail the limits of applicability or counterexamples.

\section{RESEARCH METHOD}l

This study intends to explore the difficulties encountered by students
enrolled in the General Physics courses when it comes to understand phenomena involving hydrodynamics concepts.
The first tool used for such a purpose consisted of a careful analysis of errors in the tests of the Physics 2 
students at the Faculty of Engineering of the University of the Republic, in Montevideo, Uruguay. This course lasted 15
weeks with a workload of five hours per week. In the course three weeks of classes were dedicated to topics of 
of fluid statics and dynamics.
The course bibliography consists of the usual textbooks \cite{halliday2001physics,tipler2007physics,young2008sears}.

We evaluated 600 exams, which corresponded to four midterm tests and two final exams
(100 answers were chosen randomly in each instance) 
\footnote{The tests analyzed correspond to the first midterm tests conducted in May 2014, October 2014, May 2015 and April 2016 
and the exams carried out in February 2016 and July 2016. 
The indicated evaluations can be accessed by following the link: https://eva.fing.edu.uy/mod/folder/view.php?id=37381.}.
The tests analyzed were part of the normal course evaluation. 
In each test, one of the problem-solving questions was related to fluids, with emphasis on continuity and Bernoulli’s equations. 
For each problem, which in turn was divided into several sections, the students had to specify the calculations used 
to reach the solution. Written responses were, subsequently, reviewed one-by-one and wrong answers were classified into several categories, 
according to the most frequent errors.  
Analyzing these errors, we formulated hypotheses about the students' conceptions that reflect those wrong interpretations. 

Next, to validate these hypotheses, we designed three new problem scenarios (Appendix A).
These were raised in interviews carried out with 16 students enrolled in the Science and Engineering courses. 
All the interviewees met the requirement of having successfully passed a general physics  college course that covered topics 
related to ideal fluids’ hydrodynamics. We asked the interviewees to solve problems \textit{aloud}, complementing their 
oral reasoning with written diagrams. All the interviews were audio recorded.

\section{RESPONSES OF ENGINEERING STUDENTS IN WRITTEN TESTS}

We analyzed the responses given to six problems that were divided, in turn, into several parts. 
To determine the most common errors and to infer to which conceptual difficulty they were related, 
we established categories to register the type of error and the number of students making it in written tests. 
Errors that were repeated by at least 5\% of the students that took the test were considered recurrent. 
Four recurring errors were identified.

\subsection*{Error 1: The pressure of a fluid in motion is the same as the
pressure of a fluid at rest.}

\begin{figure}[h]
\includegraphics[width=0.98\columnwidth]{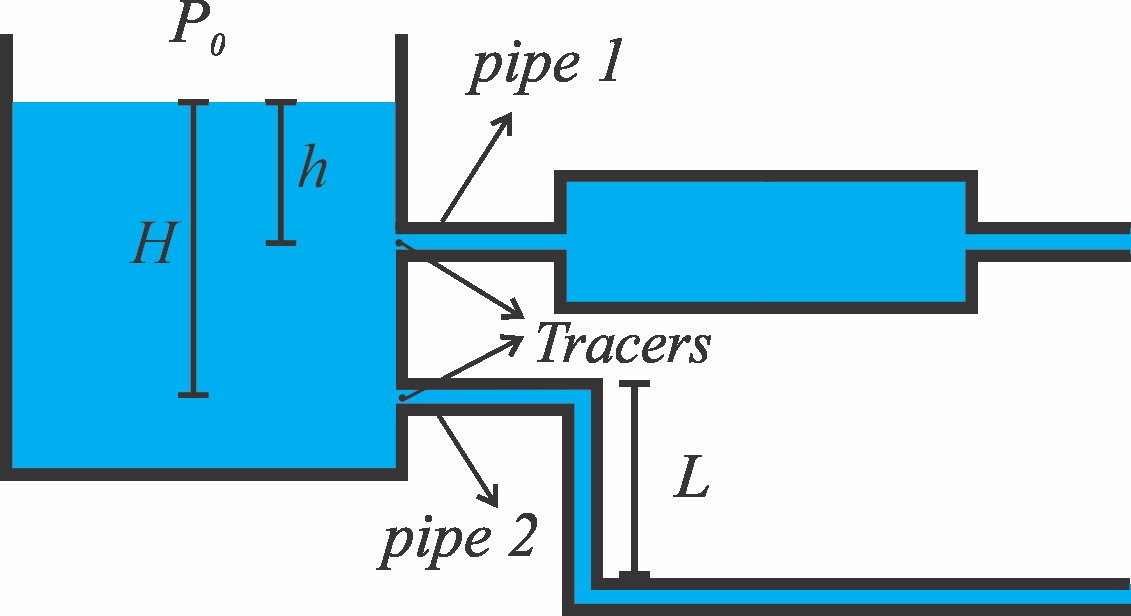}
\caption{
Container open to the atmosphere with pressure $P_0$ and  two draining pipes. 
The dimensions and the places where the tracers are released are also indicated.}
\label{fig3}
\end{figure}

To illustrate this error, Fig.~\ref{fig3} shows
the diagram of the exam carried out in February 2016. A open tank  containing water, was slowly  emptying
through pipes 1 and 2. The pressure along each horizontal pipe
should be determined according to the parameters. In the analysis of
the written tests, 27\% of the students assumed that the pressures in the first section of pipes 1 and 2 were equal to the hydrostatic, that
is, to $P_0+ \rho gh$ in pipe 1 and to $P_0+\rho gH$ in pipe 2, disregarding the velocity contributions.

It is important to highlight that in all the cases in which the pressures
were calculated as if the fluid was at rest, the continuity and
Bernoulli's equations were used to determine both the pressures and
velocities of the fluid downstream. Assuming that the pressure of a
fluid in motion is the same as that of a fluid at rest implies the
non-recognition of pressure gradients in the regions where the magnitude
of fluid velocity changes.

\subsection*{Error 2: In vertical pipes of  uniform cross-section, fluid velocity
increases due to gravitational acceleration.}

This error was observed in the  second part of the exam conducted in February 2016 which 
indicated that two tracers entered the pipes' mouths, at points 1 and 2 (indicated in Fig.~\ref{fig3}). 
Students  were asked how long the tracers \footnote{Tracers are small, colored elements that have the same density of the fluid and follow its motion without affecting its hydrodynamic 
properties.} would take to reach the end of each pipe.

Some students (9 \%)  assumed that the tracers
were accelerating along the vertical stretch of the length $L$  of pipe 2. This error was observed only in this problem. Other tests did not
ask for transit times or require analyzing the characteristics of a
moving fluid in vertical pipes. Despite the fact that only a  relatively small percentage of students assumed an accelerated motion, it was
considered that it could be  a deeply-rooted conception.

Confusing the tracer's behavior with a free-falling particle indicates that the students still considered the fluid as a set of
particles or elements that do not interact with each other. Therefore, in the context of this question, 
they neglected the principle of conservation of mass and did not recognize which forces act on a element of the fluid
flowing through a vertical pipe.

\subsection*{Error 3: For a fluid to be at rest in a vertical pipe, the pressure
difference between its extremities must be zero.}

\begin{figure}
\includegraphics[width=0.598\columnwidth]{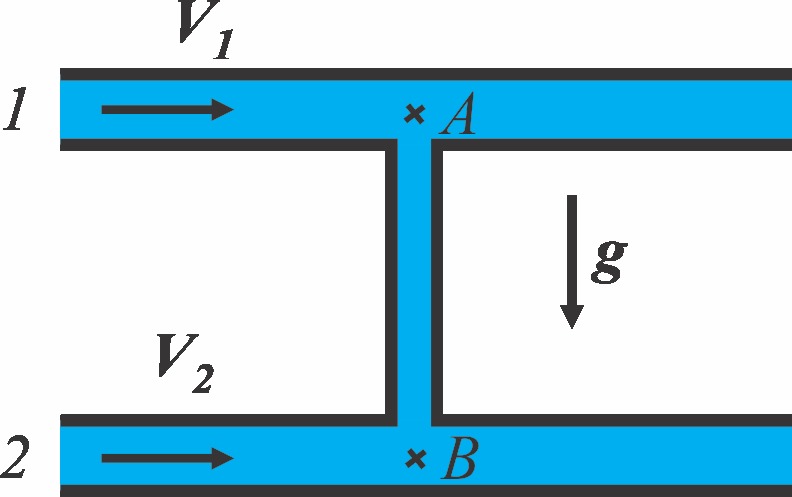}
\caption{A fluid is flowing with different velocities in two horizontal pipes joined by an another vertical pipe.
}
\label{fig5}
\end{figure}

This error was evidenced in the first midterm test carried out in May 2015.
In the hydrodynamic problem of this test, we described a situation in which two horizontal
and parallel pipes, in which fluid flowed at different speeds, were
connected by a vertical pipe, as indicated in Fig.~\ref{fig5}. The students
had to determine the pressures in points $A$ and  $B$  that would allow for the fluid
in the vertical pipe to remain at rest. When the students'
responses were evaluated, we  found that 12\% assumed that, in order for the fluid to remain at rest,  the
pressure difference between points $A$ and $B$ should be zero.

\subsection*{Error 4: Applying Bernoulli's equation to two points of a fluid,
one of which is at rest.}

\begin{figure}
\includegraphics[width=0.875\columnwidth]{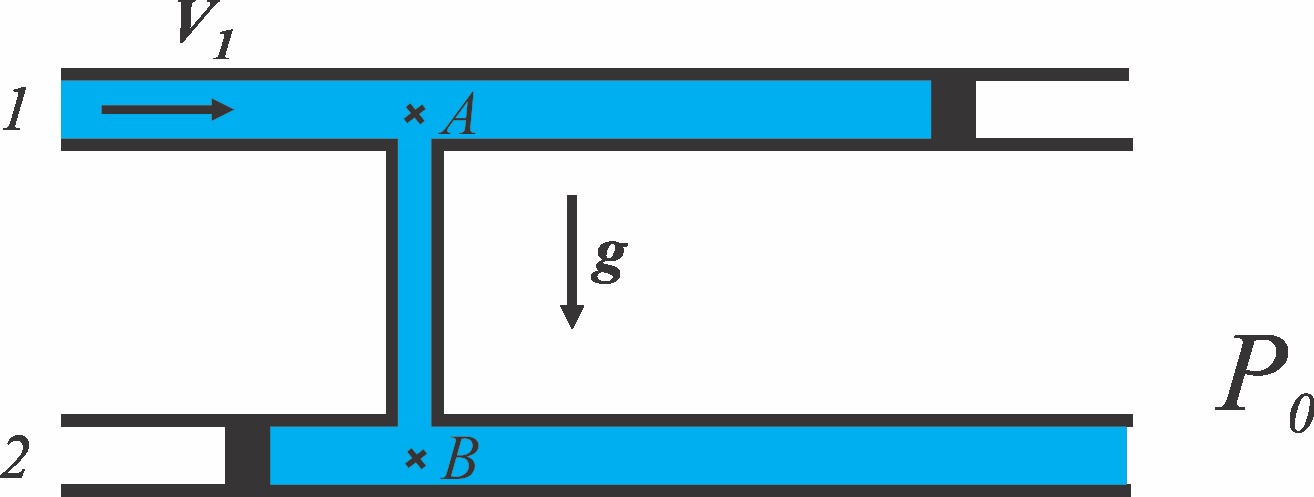}
\caption{Same setup as in Fig.~\ref{fig5} but in this situation two ends are blocked.}
\label{fig6}
\end{figure}

This error was the most commonly detected in all the tests. 
A clear example was found when analyzing the solutions provided by students in 
the second part of the problem in the midterm test carried out in May 2015 
(briefly commented in the context of Error 3). In that part of the problem, it was assumed that two plugs were placed at both pipes’ ends, 
as shown in Fig.~\ref{fig6}. In this section, the fluid enters only through the first pipe, and descends through the vertical one. 
The students were asked to determine the force exerted by the pipe walls on the plug placed at the right end. 20\% 
of the solutions presented the aforementioned error: to determine the pressure inside the plug, students applied Bernoulli’s
equation between a point at the inlet of pipe 1 and a point in the middle of the inner side of the plug. 
In this error, some students justified using Bernoulli’s equation by arguing that: 
\textit{``fluid velocity decreases down to zero at the point where the plug is.''}

\section{INTERVIEWS}
\label{sec:int}

The analysis of written exams led to a series of questions in relation to the understanding of hydrodynamic concepts. 
In order to find out how students link changes in velocity to changes in pressure, if they recognize the origin of the 
forces exerted on a fluid element and how these are linked to its motion, sixteen interviews were conducted with students 
attending Science and Engineering courses. 

These interviews posed three problems that were specifically developed for this purpose. 
The questions, shown in the Appendix,  were formulated one by one, following the  order indicated. 
The problems required the students to qualitatively solve the problems,
“\textit{thinking out loud}” their answers and, if they considered it appropriate, they were free to go back and rectify any of their
previous answers.  Continuity and Bernoulli’s equations were not mentioned before the interviews. The interviews were audio recorded.

The next section indicates both the concept  (\textit{Cx.x}) that students were expected to apply and the errors (\textit{Ex.x}) they made.

\subsection*{Problem 1}

\subsubsection*{Question A: Movement of the fluid in a vertical pipe}

\begin{figure}
\includegraphics[width=0.6\columnwidth]{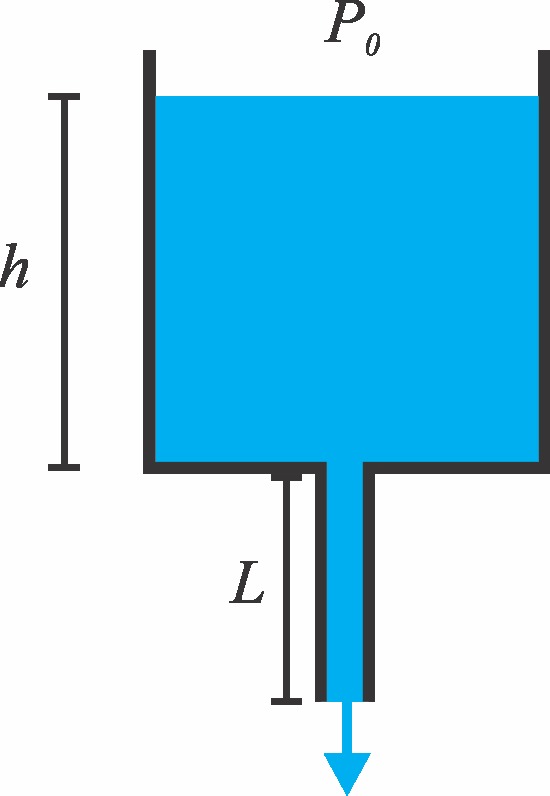}
\caption{Open container with constant level $h$ discharging through a vertical pipe  of length $L$.}
\label{fig7}
\end{figure}

The first situation considered a container with water up to $h$ height,
which was kept constant. The outlet mouth of the container, located at
the bottom, was connected to a vertical pipe of length $L$ and uniform
section, as shown in Fig.~\ref{fig7}. It was explained to the students that
the system was immersed in an environment maintained at atmospheric pressure.

First, the students were asked to describe the motion of a volume element of the fluid when descending down the pipe.
This question was posed with the objective of linking the flow of a volume element to the pressure and the forces acting on it.

\begin{itemize}
 \item 
C1.1: From the continuity equation, we concluded that the velocity of a
fluid element is constant when descending through a vertical pipe of
uniform section.

\item E1.1: The fluid makes an accelerated movement.    

\end{itemize}

 
In the interviews, 50\% of the students considered that the fluid made an accelerated movement (E1.1). 
The arguments that explicitly justify this statement can be divided into three groups. The responses 
of the first group were based on energy conservation. The students argued that, as the fluid descends, 
its gravitational potential energy decreases and, therefore, its kinetic energy increases. 
An example of this type of response was: \textit{``When falling (the fluid element), the gravitational
potential energy is converted into kinetic energy, thus increasing velocity.''}

Within this group, two students noted that matter should be conserved in the process. However, 
they erroneously considered that when the fluid velocity increased, its density decreased.
Thus, one student argued that inside the pipe, \textit{``the flow may not change... Therefore, 
the only thing that could be happening to increase velocity is a greater dispersion of water particles below, 
which diminishes the water density at the pipe’s outlet.''}

The second group of students took into account the pressures along the pipe, noting that the pressure at a certain point is 
due to the weight of the upstream fluid column. Therefore, if pressure increases with depth, velocity must also increase. 
A student from to this group explained: \textit{``The water column becomes bigger as depth increases. Since this pushes it down, 
the fluid element should accelerate.''}

Finally, the third group, which, in this case, comprised by a single student, invoked Bernoulli’s equation,
without taking into account the term of pressure. This student pointed out that \textit{``...velocity must increase with depth. 
Thus, a fluid element must accelerate when descending...''}

\subsubsection*{Question B: Differences of pressures}

\begin{figure}[h]
\includegraphics[width=0.6\columnwidth]{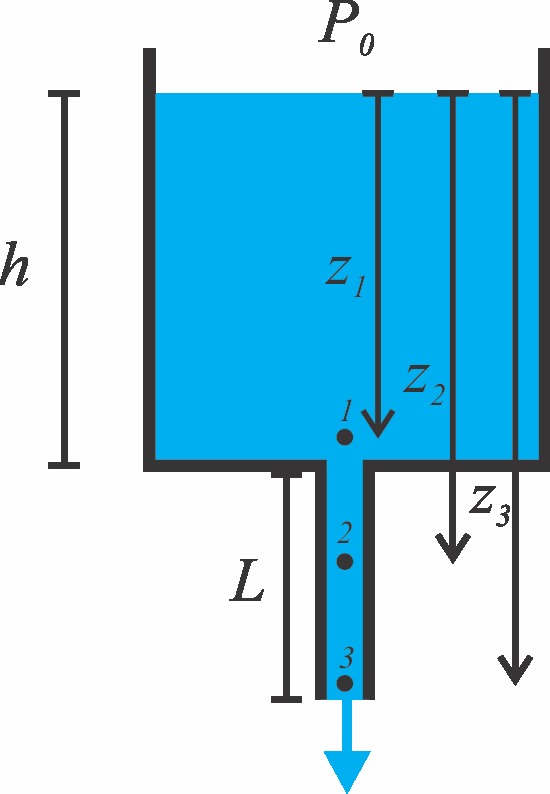}
\caption{Open container with a vertical draining pipe. Three points, $1$, $2$, and $3$ at different 
heights $z_1$, $z_2$, and $z_3$, are indicated.}
\label{fig8}
\end{figure}

Next, the students were asked to compare the differences of pressures at different points of
the system described above with those of a fluid if there were a plug at the pipe’s end. 
Thus, students were required to compare  hydrostatic and hydrodynamic pressures
in similar configurations). 
Three points were considered: one above the container’s mouth (point 1) and two 
inside the vertical pipe (points 2 and 3), as indicated in Fig.~\ref{fig8}.

\begin{itemize}
 \item 
C1.2: When the fluid is in motion, the difference of pressures between
points 1 and 2 is lower than the difference of hydrostatic
pressures $ \rho g(z_2-z_1).$ This is because the fluid increases its velocity
once it enters the vertical pipe.

\item E1.2: When the fluid is in motion, the difference of pressures between
points 1 and 2 is equal to the hydrostatic pressures $ \rho g(z_2-z_1).$
\end{itemize}

Approximately, one-third of the students indicated E1.2 arguing
that, \textit{``...the column of fluid above these points is the same in both
cases.''}

\begin{itemize}
 \item  C1.3: When the fluid is in motion, the difference of pressures between
points 2 and 3 is equal to the hydrostatic pressures $ \rho g(z_3-z_2)$
since no velocity changes are observed in the pipes in the uniform
section.

\item E1.3: When the fluid is in motion, the difference of pressures between
points 2 and 3 is lower than the hydrostatic pressures $ \rho g(z_3-z_2).$
\end{itemize}

There were two types of responses associated with E1.3. Two students assumed, correctly, 
that the fluid was moving at constant velocity (C1.1) inside the vertical pipe. They concluded that the difference of 
pressures should be zero, since the net force on a volume element was also zero. Thus, they ignored gravitational force. 
One student in particular justified the response as follows:
\textit{``The pressures in the upper and lower parts of the container are the same. Thus, when the fluid is moving at constant velocity, 
the pressures are equal along the pipe.''}

On the other hand, those who considered that the fluid was accelerating assumed that the difference of pressures decreased as
a consequence of Bernoulli’s equation, based on the idea that the theorem states that a higher velocity implies lower pressure. 
This is in agreement with the following argument: \textit{``The pressure in points 2 and 3 changes when compared to the hydrostatic pressures 
(because the fluid is in motion). At point 3, the fluid flows faster, so the pressure difference is lower.''}

\subsubsection*{Question C: Draining time}

\begin{figure}[h]
\includegraphics[width=0.9\columnwidth]{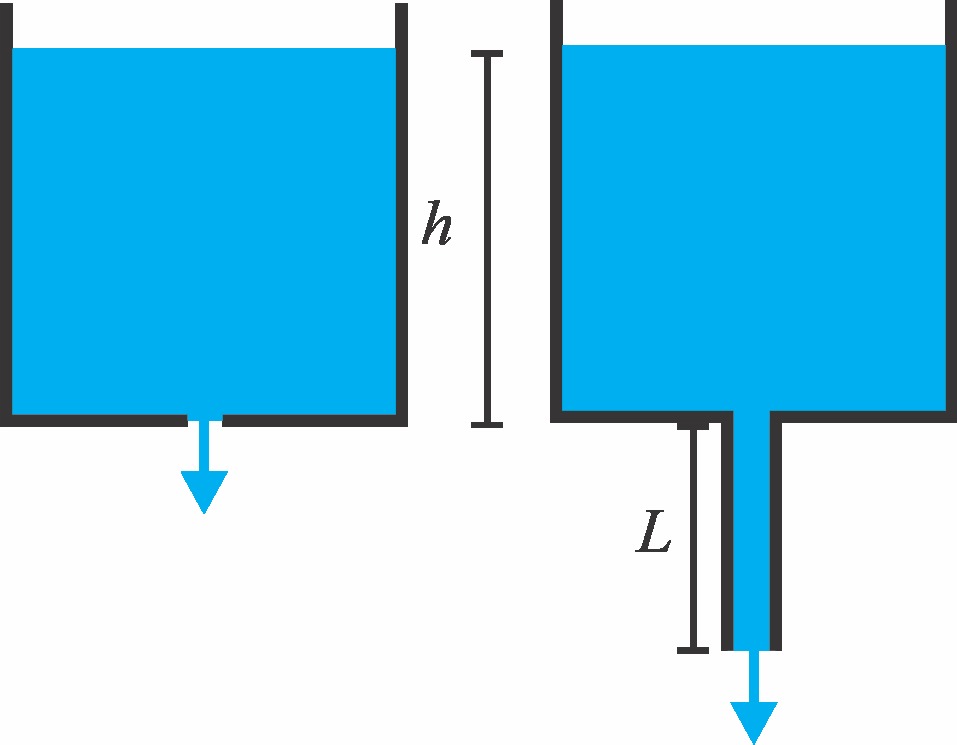}
\caption{Two similar open containers, one of them with a vertical draining pipe of length $L$. The initial water level is 
$h$ in both containers.} 
\label{fig9}
\end{figure}

The last part of the first problem required students to compare the draining times of the containers mentioned in previous sections, 
with one identical container (Fig.~\ref{fig9}) without the vertical pipe. This question was posed to find out the variables influencing 
the exit velocity of the fluid in a container,
according to the students. This question also tried to assess their ability to link the velocity to the work done by the different pressures. 

\begin{itemize}
 \item  C1.4: The draining time of a container with a hole that opens to the
atmosphere is greater than that of the container that was emptied through a
vertical pipe.

\item E1.4: Both systems have the same draining time, since they have the
same output velocity.
\end{itemize}

Approximately two-thirds of the students responded that the draining
times were the same (E1.4). The most frequent argument between the
students of group C1.1 (constant velocity inside the pipe) was that
the amount of water above the mouth of the base of each container was
the same. This implied the same output velocity and, therefore, the
same draining time. One response was: \textit{``They (the containers) drain
at the same time, since they have holes with the same area and have the
same output velocities. This is due to the fact that the water
pressure above is the same.''}

Responses incurring in E1.4, were also given by some students who considered
that the fluid accelerated inside the vertical pipe (E1.1). Other
students belonging to the group that gave the E1.1 response also
included the E1.4 argument. However, they indicated that the flow at
the containers' mouths (one open to the atmosphere and another with
water draining through the pipe) was the same.

\subsection*{Problem 2}

\begin{figure}
\includegraphics[width=0.7\columnwidth]{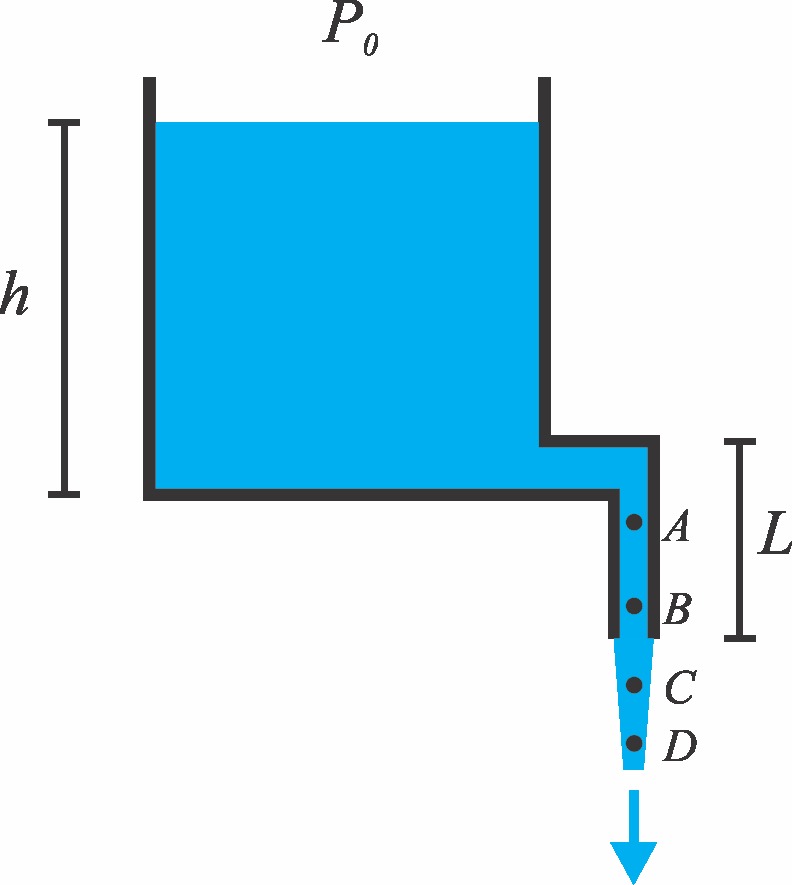}
\caption{Open container with water kept at a constant level $h$. The fluid drains through
 a pipe which ends with a vertical section of length $L$. Four points, $A$, $B$, $C$, and $D$ are represented, two inside the pipe
 and the other two  after the mouth of the pipe.
}
\label{fig10}
\end{figure}

The second problem referred to a tank containing water at a constant $h$ level and with a pipe of vertical length $L$ 
at its lower right end, as shown in Fig.~\ref{fig10}. The shape of the water jet after coming out of the pipe was also shown.

\subsubsection*{Question A: Comparing velocities and pressures}

The first question considered four points, two inside the vertical
pipe ($A$ and $B$) and two ($C$ and $D$) outside. The students were asked to
compare velocities and pressures of the fluid at these four points.

The goal of pointing points $A$ and $B$ again as points of interest was to assess if 
students could answer consistently when asked about velocities and pressures inside the vertical pipe,
in a context slightly different from the one presented in the first problem. 
Changing the context did not alter their answers about velocity and pressure.

Regarding the velocities and pressures of the jet flow, we expected them to answer:

\begin{itemize}
 \item C2.1: The velocity of the fluid at point $C$ is greater than inside
the pipe, and this is even higher at point $D$.

\item C2.2: The pressure at $C$ and $D$ is atmospheric. 
\end{itemize}

The errors found were the following:

\begin{itemize}
 \item  E2.2a: The pressure at $D$ is lower than the pressure at $C$, since the
higher the velocity is the lower the pressure will be.

\item E2.2b: The pressure at $D$ is higher than the pressure at $C$ due to the
extra pressure exerted by the fluid column above point $D$.

\end{itemize}

With regard to velocities, all the students answered C2.1. However, concerning the pressures at
points $C$ and $D$, half of the students answered E2.2. Approximately,
one-third of these students replied with E2.2b, while the rest were based on E2.2a.

Two students’ responses, consistent with E2.2a and E2.2b, were 
\textit{``The pressure at point $D$ is lower than that of point $C$. Presumably, 
this is because of Bernoulli, as the higher the velocity is, the lower the pressure will be. 
This point ($D$) presents a higher velocity and lower pressure...''} and the other:
\textit{``At point $D$, the water column and its velocity are higher. 
Thus, the water column exerts a net force downward, so the pressure at $D$ must be higher than in $C$.''}

\subsubsection*{Question B: Diameter of the Jet}

Next, the focus was shifted to the decreased diameter of the water jet section as the fluid flowed down. 
Thus, the students were asked the cause of that reduction. 

The objective was to identify how students applied the continuity equation to contexts different than usual,
namely, a situation where the fluid was not confined to a pipe. Furthermore, we sought to analyze how students managed 
to generate a coherent model to explain this phenomenon that could also be compatible with their previous responses on water 
jet velocity and pressure at different points. 
We expected, therefore, answers that linked the pressure at different jet points to velocity and the continuity equation. 
Next, we focused on the decreased diameter of the water jet section as the fluid flows down. Thus, the students were asked the cause of that
reduction.

\begin{itemize}
 \item  C2.3: The continuity equation indicates that the jet section at point $C$ (lower velocity) is greater than at point $D$ (higher velocity).

\item E2.3: The atmospheric pressure increasingly “compresses” the water jet. 
\end{itemize}

Twenty-five percent of the students answered E2.3. The main argument was that when the jet velocity increased, its pressure decreased. 
Thus, the difference of pressures between the water and the atmosphere was responsible for the effect observed.

In agreement with these arguments, but without using the concept “greater velocity-lower pressure,” a student concluded the following: 
“\textit{The pressure at point $C$ is greater than at D, since the water is more compressed at D. The water pressure decreases since it
is more compressed by the air. As the water pressure decreases, the difference in pressures between air and water is greater at $D$ than at C.}” 

\subsubsection*{Question C: Forces acting on a fluid element}

\begin{figure}
\includegraphics[width=0.25\columnwidth]{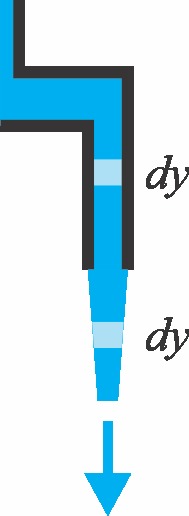}
\caption{Magnification of Fig.~\ref{fig10} in which two elements of the fluid with length $dy$ were represented.
}
\label{fig11}
\end{figure}

The final question presented a volume element of the fluid inside the vertical pipe and another outside of it, as indicated in Fig.~\ref{fig11}.
This scenario required students to describe which forces acted on each fluid element and how they are related to each other.

This question aimed to find out how students considered the effect of atmospheric pressure and how different parts of a moving 
fluid interact with one another. 
The objective was also to analyze the compatibility of these answers with the way in which students understood the movement of the fluid. 

Firstly, we considered  the forces acting on a  \textit{confined fluid element}. The correct and incorrect responses
analyzed were the following:

\begin{itemize}
 \item  C2.4: The volume element inside the pipe is subjected to forces
exerted by the action of pressures on the surfaces and its weight. The
net force is zero.
 \item  E2.4: Incorrect analysis of forces on the element with possible
contradictions in the resulting motion.
\end{itemize}

As for the volume element in the pipe, two students considered that the water was moving at a constant velocity and argued that this type
of movement implied constant pressures along the pipe, which meant that the only acting force was weight. 
When asked to reconcile this answer with the notion that the net force on the fluid element was zero, 
they recognized there was something incorrect in their reasoning. 
However, they were not able to find the source of this contradiction. 

Half of the students made reference to a weight force and a downward force exerted by the fluid on the volume element (E2.4),
regardless of whether they assumed that the fluid was moving at a constant or accelerated velocity. 
When questioned about the origin of this force, the most common argument was “\textit{due to the weight of the water column.}”
Regarding the upward force exerted by the fluid below the volume element, 25\% of the students considered that, since all the fluid was flowing down, there was no reason 
to take into account that force.

When relating the movement of a fluid element to the sum of the forces acting on it, 
the students mainly provided two intuitive Aristotelian responses, associated with the necessary condition of 
the water to come out of the pipe. One example of this is evident from the following response: 
“\textit{The weight and upstream pressure is greater than that (pressure) downstream. This is why the water comes out of the pipe.}” 
This response also shows the existence of a language (often used by students), in which pressure is equal to force.

Next, we considered an \textit{unconfined fluid element} as follows:

\begin{itemize}
 \item C2.5: The volume element in the jet flow is at atmospheric pressure and, as such, weight is the net force acting on it.
\item E2.5: Incorrect analysis of forces acting on the element with emphasis on the force exerted by the upstream liquid. 
\end{itemize}

Twenty-five percent of the students provided answers based on E2.5, since they considered that the part of
the fluid below the volume element was not exerting any force, as all fluid was “falling.” A common answer to this question was the following: 
“\textit{The net force is exerted downwards. The force due to the column above is the one causing the acceleration.}” 
In this particular answer, the student did not refer to the weight. However, there is still the concept regarding the effect of 
the fluid above the volume element.
 
It is worth noting that a couple of students had difficulty interpreting the net force due to the atmospheric pressure, 
since they could not recognize it was equal to zero.

\subsection*{Problem 3}

\begin{figure}
\includegraphics[width=0.42\columnwidth]{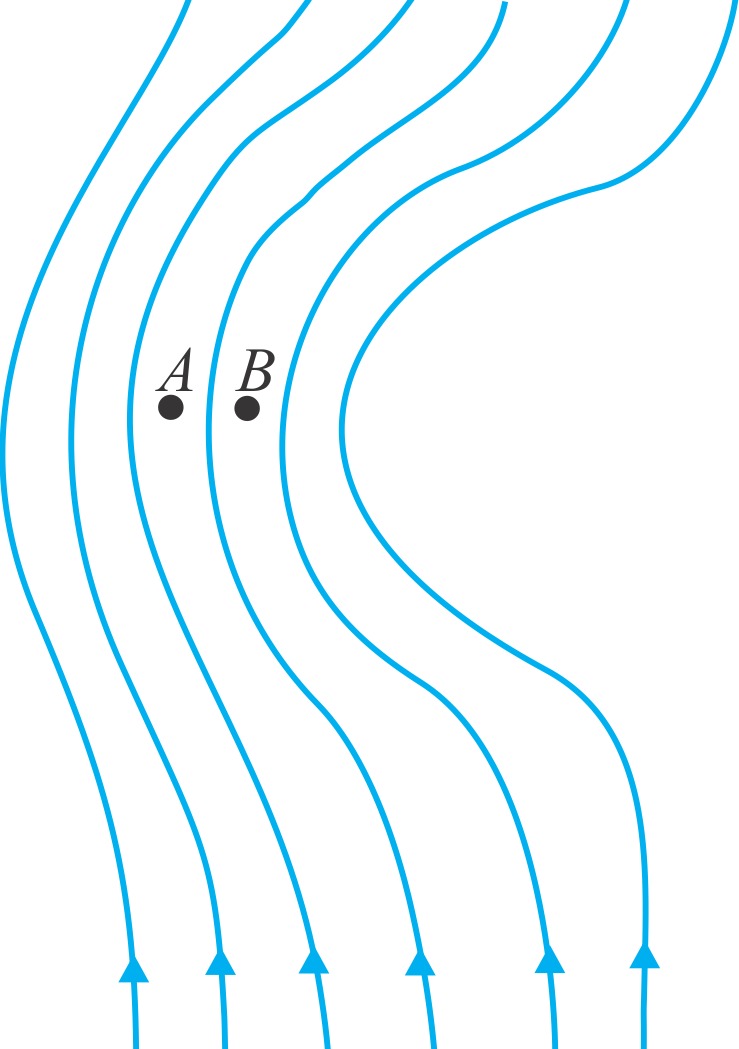}
\caption{Streamlines of a fluid moving in a vertical direction
which encounters an obstacle (not represented here).
}
\label{fig12}
\end{figure}

The last problem of these interviews presented a confined flow moving vertically at high speed. 
The current lines were strongly deformed, when encountering an obstacle. Figure~\ref{fig12} shows this situation 
(the diagram does not show the obstacle, since it’s not of interest). 

The problem required students to compare pressures at the points $A$ and $B$, which are marked in the figure. 
To the students, it was explained that the velocities of the flow particles passing through those points were equal
and that the points were at the same height. This problem was posed to once again assess which variables the students 
considered the pressure depended on at a point of the fluid in motion. 

\begin{itemize}
 \item C3.1: The radius of the current curvature lines is associated with pressure gradients acting normally to their direction.
The pressure at point $A$ is higher than that at point $B$.

\item E3.1a: Equal height implies equal pressures at points $A$ and $B$.

\item E3.1b: Equal velocity and height imply equal pressures at points $A$ and $B$.
\end{itemize}

All students answered with E3.1, explaining the two arguments. Those presenting a simpler view of the problem (two interviewees) argued 
that the pressures at points A and $B$ should be the same, given that “\textit{those points are of equal height.}” In this case, the idea was that, 
if the height is the same, the pressure must also be the same, regardless of fluid motion (E3.1a). 
The remaining interviewees considered a possible dependence of pressure from velocity (E3.1b). 
An example of this type of response was: “\textit{A greater concentration of lines corresponds to higher velocity. 
The pressure depends on the height and, if the velocity is the same, the pressures are the same.}”

\section{DISCUSSION OF THE RESULTS AND STUDENTS' MISCONCEPTIONS}

After analyzing the responses provided by the students in written exams
and interviews, it was possible to infer the existence of a number of
misconceptions in the context of hydrodynamics of ideal fluids. This
section includes a description of each.

\subsection{Transition from hydrostatic to hydrodynamics}

The analysis of written exams highlighted that many students consider that the pressure of a fluid in
motion is the same of the fluid at rest. In addition, from the interviews we conclude that many students maintain
the misconception that the pressure is associated only to the weight of the liquid column that the fluid has on top. 
This misconception, which had already been reported in the literature \cite{besson2004students,goszewski2013exploring,loverude2010identifying}
in the context of hydrostatics, appeared transversally in the various situations presented in the interviews. In particular, 
it was clear from the responses provided in Problem 1 about the analysis of the pressures’ difference between a point above 
the container’s hole and the point inside the pipe. Similarly, this was also evident in Problem 2,
in which the pressures corresponding to points $C$ and $D$ were compared. Those with this deeply-rooted concept tend to apply it to hydrodynamics, 
since they are not able to provide another reason for which the pressure could change at a certain point of the fluid.

\subsection{Compressed and expanded fluids}

The basic characteristics of liquids and gases are studied in high school Physics and Chemistry courses. 
One of the main differences between these matter states is that, under usual conditions, 
liquids like water are not compressible. This difference, which a priori is simple, is not always present in
the models used by the students to analyze fluids in motion. This was already reported in the literature, Brown \textit{et al.} \cite{brown2017engineering}
demonstrated how students believe that liquids are compressed when flowing through a narrow pipe. 
This conception was present during the interviews, and it is mentioned in the explanations provided for Problem 2, 
in which we asked the students why the water jet was narrower when exiting the pipe. In this situation, 
several students associated the narrow jet with water compression, without taking into account that the 
liquid is an incompressible fluid. It is important to note that although this error was already present
in the literature, the context presented in this study was different, since the fluid was not confined to a pipe.
It is noteworthy, then, that this deeply rooted conception was present. 

We also noted a misconception not reported so far regarding how liquids descending a pipe could expand. 
This idea was present in the answers of two students, who tried to fit their model of an accelerated fluid along
the pipe, presented in Problem 1, with the principle of conservation of matter. This led them to consider that water density must decrease with depth.

\subsection{Higher velocity  always means lower pressure}

The error that the pressure always decreases as fast as a fluid moves comes from oversimplifying the results 
of Bernoulli’s equation. This is in agreement with some justifications provided by the students, describing how 
the pressures were along the falling water jet, presented in Problem 2, as they assumed that if the fluid moved increasingly faster, 
its pressure must be lower, ignoring the fact that the fluid should be at atmospheric pressure.

In Problem 3, the students had to compare the pressures between points $A$ and $B$. Some of them took into account that the pressure
could be dependent on velocity. When recognizing that their magnitudes were the same in both points, the students discarded automatically 
the possibility that different pressures were present in $A$ and $B$. Thus, the concept that the velocity‘s magnitude defined the relationships 
between pressures prevailed, ignoring a potential incidence of the curvature of the current lines on the existence of a pressure gradient on
the normal direction to the lines.

\subsection{The force on a fluid element is exerted by the fluid above it}

The students considered only the force exerted by the upstream fluid. This response was common and used to justify a possible accelerated
movement of the fluid along the vertical pipe, presented in Problem 1. However, they forget that the volume element considered also exerts
force downwards against the fluid to be displaced and that by Newton's third law, an upwards force must be included in the balance of 
forces acting on the fluid element. This misconception, not reported in the literature in a hydrodynamic context, was also detected in
the answers given by the students in Problem 2, where they had to analyze forces acting on a volume element. In that problem, 
not only it seemed reasonable to them that the upstream fluid was the only one exerting a force, but they also considered this 
could work for both confined and non-confined fluids. A similar difficulty to describe the interaction between fluids and their
related forces and pressures in the context of hydrostatics was reported by Loverude  \textit{et al.} \cite{loverude2010identifying}.

\subsection{Naive interpretations of Bernoulli's equation}

As previously mentioned in the article, Bernoulli’s equation links the kinetic and gravitational potential energies per 
volume unit with the work performed per unit volume by pressures at the borders of a system. However, when explaining the 
type of movement performed by a volume element of the fluid when descending a vertical pipe, as presented in Problem 1, 
the students took into account only the terms of gravitational potential energy and kinetic energy. Then, they are																																																																																																																																																				 ignoring arguments linked 
to the work by the pressures on the borders of the volume element, as if the behavior of the fluid were similar to that of a 
set of particles or elements not interacting with each other. The error regarding the behavior of a fluid in motion had not been previously reported.

Similarly, the clarifications associated with the discharge times of containers presented in Problem 1 also ignore the work 
performed by the pressures. The error that led students to consider the containers were draining at the same time implied that
the water velocity at the mouth of each container base was the same. This reasoning relies exclusively on the fact that the 
initial water levels were the same in both containers, and deeming irrelevant the fact that the pressure of the container 
mouth at the pipe inlet was different from the atmospheric pressure. This lack of consideration of different pressures was 
reported by Vega-Calderón \textit{et al.} \cite{vega2017dificultades}.
 
The analysis of the written exams highlighted the difficulties of the students to understand that Bernoulli’s equation 
is applied between two points of the same streamline or stream pipe. Assuming that it can be applied to any two points 
of a fluid appears to be a naive interpretation of the equation. This concept might derive from the first applications of 
Bernoulli’s equation shown in textbooks  \cite{halliday2001physics,young2008sears,serway2013physics,tipler2007physics}.
These books show that, if the fluid is at rest, the hydrostatic expression is recovered. The student could then quickly infer that
Bernoulli’s equation might also predict a correct result when at one of the points the velocity is equal to zero. 

\section{CONCLUSIONS}

Based on the analyses of midterm tests and exams, and the results of interviews carried out with students who had passed a general 
physics course that covered topics of fluid mechanics, we analyzed the most common conceptual difficulties related to ideal fluid hydrodynamics. 
We found various misconceptions, some of them  already described in the literature, and others not yet reported. 
Among the latter, we observed that many students faced difficulties in recognizing how the volume element of a fluid in motion interacts with 
its environment. In several cases, they assumed that the behavior of a confined fluid is similar to that of a set of particles or elements that 
do not interact with each other; while in other cases, a volume element in motion is affected only by the upstream fluid. 

Another novel aspect to highlight is that many students applied Bernoulli’s equation to different points of the fluid without taking into
account that some are in motion and others are at rest. They fail to consider that this equation is derived from a conservation law, applied
to a volume element moving along a streamline or belonging to the same stream pipe.

Many students also assumed that hydrostatic conditions determine the pressures of a fluid in, confined or non-confined, motion. 
The belief that pressures in  moving fluids  are described in the same way as  hydrostatic pressures could be dependent on the particular context.
We observed that this is a strongly rooted idea among students who answered the questions in which the fluid moves through a vertical pipe. 
This misconception was not as common among students that solved the problems in which the fluid moves inside 
horizontal pipes that change section. 
However, in this case, the students could have applied correctly the continuity and Bernoulli’s equations, without realizing that a 
pressure gradient is a factor associated to velocity changes in space.

There seems to be common elements in the conceptual difficulties observed. The students faced difficulties to understand how different parts of 
a fluid in motion interact, failing to link kinematics with dynamics. Specifically, students are not able to link pressure gradients to forces acting 
on a fluid element and its changes in velocity. Perhaps, the most obvious case of this missed association is when comparing the pressures between 
two points at equal height, within a flow in which the current lines are curved. In this case, the students were not able to recognize that the
fluid elements are accelerating and, accordingly, there must be a pressure gradient.

How hydrodynamic contents are presented in textbooks could be one of the causes responsible for many of the conceptual difficulties encountered. 
In the textbooks of General Physics \cite{halliday2001physics,young2008sears,serway2013physics,tipler2007physics}, 
the study of ideal fluids hydrodynamics is practically limited to applications associated with the continuity and Bernoulli’s equations.  
These contents are treated with the results derived from the work and energy theorem and  the law of conservation of mass.
Thus, dynamic analysis of forces operating on a fluid element in motion and the causes for which such element moves with uniform or non-uniform
velocities are excluded or superficially carried out, which results in a partial understanding of these phenomena. 

One of the limitations of this study is the small number of conducted interviews. However, these interviews offered us a thorough
picture of possible interpretations provided by the students  and supported by the analysis of hundreds of exams and tests, which revealed
the aforementioned errors, which were then confirmed and examined in-depth during the interviews. A different type of student population
would need to be analyzed, possibly from other countries, although we would expect to find students with the same type of misconceptions,
since the strategies and textbooks that the students in this study used are employed internationally.

The results of this study highlight the need to deepen on students’ conceptual understanding and how to connect Newtonian mechanics with 
other branches of Physics. Our results suggest that a poor conceptual understanding of dynamics is the main cause for the numerous conceptual
difficulties encountered. We also believe that the errors found are fundamental to reconsider how ideal hydrodynamic fluids contents are presented
in General Physics courses. The knowledge of the causes that prevent students from appropriating hydrodynamic concepts is crucial to develop instructional materials and standardized tools.
We  are convinced that understanding students' misconceptions and difficulties is a necessary previous step
to improve undergraduate curriculum in Sciences and Engineering.

\begin{acknowledgments}
The authors would like to thank financial support from the \textit{Programa de Desarrollo de las Ciencias B\'asicas} (MEC-UdelaR, Uruguay).
\end{acknowledgments}

\appendix* \section{Problems posed during the interviews}

\subsection*{Problem 1}

A container is filled with water up to h height as shown in the Fig.~\ref{fig7}. A draining pipe of
length $L$ (of the same diameter as the outlet) is located at the outlet mouth.  Suppose that through a device, not indicated in the figure,
the water level of the container is maintained at constant height $h$. 
Assume that the viscosity of the water is not significant.  

\textbf{Question A:} If $v$ is the velocity of a fluid element entering the inlet of the draining
pipe of length $L$, what happens to the velocity of the fluid element flowing down the pipe?  

\textbf{Question B:}  Next, consider points 1, 2, and 3, which are marked in the diagram enclosed (Fig.~\ref{fig8}).  

\begin{enumerate}[label=(\roman*)]
 \item  Is the difference of pressures $\Delta P$ between points 1 and 2, higher, lower, or equal to $\rho g(z_2-z_1)$?  
\item  Is the difference of pressures $\Delta P$ between points 2 and 3, higher, lower, or equal to $ \rho g(z_3-z_3)$?  
\end{enumerate}

\textbf{Question C:} Suppose that we place one identical container next to the container with the pipe 
considered in the previous questions as indicated in Fig.~\ref{fig9}.
The only difference between the containers is that the new container does not have a draining pipe.
Initially, both  containers have  an equal amount of water.
If  you remove the system that kept the water level constant in the containers, 
which container will drain first, or will both containers drain at the same time?

\subsection*{Problem 2} 

At the outlet of a tank containing water at a constant $h$ level (maintained through a system that is not indicated 
in the figure), we
connect a pipe. The water flowing out the tank passes through the
vertical section of a pipe long $L$.  As we move away, the water jet exiting the pipeline becomes narrower as it 
falls (Fig.~\ref{fig10} represents the situation described).  
Assume that the viscosity of the water is not significant.  

\textbf{Question A:} Consider points $A$ and $B$ (inside the vertical pipe), and $C$ and D, outside the pipe
as indicated in Fig. \ref{fig11}.
\begin{enumerate}[label=(\roman*)]
 \item  What are the fluid velocities at these points?  
\item What are the pressures at these four points?
\item Why does the diameter of the water jet decrease as the fluid drains out? 
\end{enumerate}

\textbf{Question B:}  Consider a fluid element of length $dy$ \textit{inside} a vertical pipe. 

\begin{enumerate}[label=(\roman*)]

\item What are the forces that act on this element? How do they relate to each other? 

\end{enumerate}

\textbf{Question C:} Consider a fluid element of length $dy$ \textit{outside} the vertical pipe.

\begin{enumerate}[label=(\roman*)]

\item Which forces act on this element? How do they relate to each other?

\end{enumerate}

\subsection*{Problem 3}  A fluid, which is moving vertically at a high velocity, encounters
an obstacle and  its streamlines deform  as shown in Fig.~\ref{fig12}.  (The obstacle is not represented in
the diagram, since it is not of our interest).  In the area where
points $A$ and $B$ are marked, the current lines make circumferential
arcs with approximately the same velocity.  Assuming that points $A$
and $B$ are at the same height in relation to the sea level, compare the
pressures between these points.

\bibliography{/home/arturo/Dropbox/bibtex/mybib}

\end{document}